\documentclass[12pt]{article}
\usepackage{graphicx}

\newcommand\pubnumber{}
\newcommand\pubdate{\today}

\def\support{\footnote{Work supported by JSPS KAKENHI Grant Number 23224007.}}

\def\Title#1{\begin{center} {\Large #1 } \end{center}}
\def\Author#1{\begin{center}{ \sc #1} \end{center}}
\def\Address#1{\begin{center}{ \it #1} \end{center}}

\newcommand\pubblock{\rightline{\begin{tabular}{l} \pubnumber\\
         \pubdate  \end{tabular}}}
\newenvironment{Abstract}{\begin{quotation}  }{\end{quotation}}
\newenvironment{Presented}{\begin{quotation} \begin{center} 
             PRESENTED AT\end{center}\bigskip 
      \begin{center}\begin{large}}{\end{large}\end{center} \end{quotation}}
\def\Acknowledgements{\bigskip  \bigskip \begin{center} \begin{large}
             \bf ACKNOWLEDGEMENTS \end{large}\end{center}}

\newcommand{\CP}{$CP$}
\newcommand{\ko}{K^0}
\newcommand{\kobar}{\overline{K}^0}

\newcommand{\klpiopio}{K_L \to \pi^0\pi^0}

\newcommand{\epoe}{\epsilon'/\epsilon}
\newcommand{\reepoe}{Re(\epoe)}
\newcommand{\kpinn}{K \to \pi \nu \overline{\nu}}
\newcommand{\klpionn}{K_L \to \pi^0 \nu \overline{\nu}}
\newcommand{\kppipnn}{K^+ \to \pi^+ \nu \overline{\nu}}
\newcommand{\kenu}{K^+ \to e^+ \nu}
\newcommand{\kmunu}{K^+ \to \mu^+ \nu}

\newcommand{\sm}{standard model}

\newcommand{\etal}{\textit{et al.}}

\newcommand{\prl}[4]{#1, Phys.\ Rev.\ Lett.\ {\bf #2}, #3 (#4).}
\newcommand{\prd}[4]{#1, Phys.\ Rev.\ D {\bf #2}, #3 (#4).}

\newcommand{\plb}[4]{#1, Phys.\ Lett.\ B {\bf #2}, #3 (#4).}

\begin{document}
\begin{titlepage}
\pubblock

\vfill
\Title{Current and Future Kaon Experiments}
\vfill
\Author{Taku Yamanaka\support}
\Address{Department of Physics, Osaka University\\
1-1 Machikaneyama, Toyonaka, Osaka 560-0043\\
Japan}
\vfill
\begin{Abstract}
Kaon experiments are now focusing on searching for new physics beyond the \sm.
For example, 
CERN NA62, J-PARC KOTO and J-PARC TREK-E36 experiments are starting up 
to study $\kpinn$ decay modes, a lepton flavor violation, and lepton universality.
\end{Abstract}
\vfill
\begin{Presented}
The 8th International Workshop on the CKM Unitarity Triangle (CKM 2014)\\
Vienna, Austria, September 8--12, 2014
\end{Presented}
\vfill
\end{titlepage}
\def\thefootnote{\fnsymbol{footnote}}
\setcounter{footnote}{0}

\section{Introduction}
Kaon experiments have played important roles in the developments of the \sm\ in particle physics,
such as the discovery of strangeness and the discovery of \CP\ violation in $\ko -\kobar$ mixing.
Fermilab KTeV and CERN NA48 experiments measured $\reepoe \neq 0$ and proved that 
\CP\ is violated also in the decay process itself
\cite{ktev_1999, na48_1999, ktev, na48}.
The non-zero $\reepoe$ results and the \CP\ violation found in B mesons~\cite{babar, belle} established 
the Kobayashi-Maskawa's model~\cite{km} as the source of 
\CP\ violation observed in laboratories, 
and made the model one of the fundamental pieces of the \sm.

However, the \sm\ cannot explain the \CP\ violation in the universe, \textit{i.e.} 
the large asymmetry between the existence of matter and anti-matter.
There should thus be new physics beyond the \sm\ that causes a large \CP\ violation.
This manuscript describes the current and future kaon experiments that are designed to 
search for new physics beyond the \sm.

\section{Experiments for $\kpinn$}
\begin{figure}[!htb]
	\centering
	\includegraphics[width=0.3\linewidth]{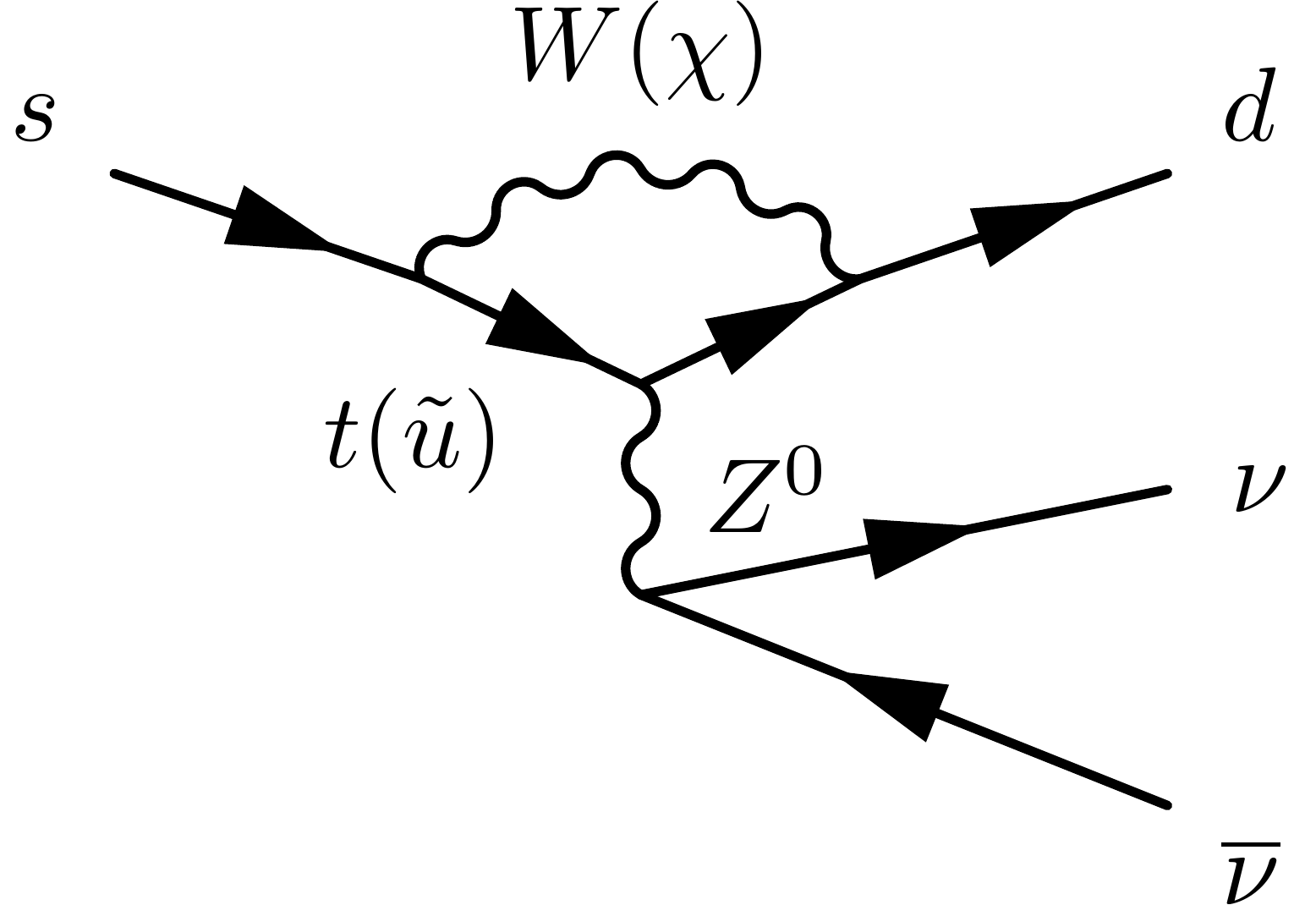}
	\caption{Penguin diagram for the $\kpinn$ decay.}
	\label{fig:penguin}
\end{figure}

To search for a small signature of new physics, backgrounds have to be small 
and its size should be known accurately.
One of such sensitive probes is the $\kpinn$ decay.
In the \sm, the decay proceeds through a penguin diagram shown in Fig.~\ref{fig:penguin}.
The \sm\ predicts the branching ratios to be small,
\(BR(\kppipnn) = 7.8 \times 10^{-11}\), and
\(BR(\klpionn) = 2.4 \times 10^{-11}\)
with $2-4$\% theoretical errors
\cite{brod_prd83_2011}.
If a new physics contributes to these decay modes by having new particles in the loop, 
it can change the branching ratios from the \sm\ predictions.
In addition, the $\klpionn$ decay mode is sensitive to new physics that violates \CP\ 
because the $K_L$ is mostly \CP-odd and $\pi^0\nu\overline{\nu}$ is \CP-even.

Experimentally, the branching ratio of $\kppipnn$ was measured 
by the BNL E787 and E949 experiments to be
\(BR(\kppipnn) = (1.73^{+1.15}_{-1.05}) \times 10^{-10}\)
based on 7 observed events by using stopped $K^+$s~\cite{bnl949_prd79_2009}.
The best upper limit on the branching ratio of $\klpionn$ was given by the KEK E391a experiment to be
\(BR(\klpionn) < 2.6 \times 10^{-8}\) (90\% CL)~\cite{e391a}.
Using an isospin rotation, the measured branching ratio of $\kppipnn$ 
gives a constraint, \(BR(\klpionn) < 1.46 \times 10^{-9}\ (1\sigma)\),
which is called a Grossman-Nir bound~\cite{grossman_nir}.

Currently, there are the CERN NA62 experiment to study the $\kppipnn$ decays and
the J-PARC KOTO experiment to search for the $\klpionn$ decays.

\subsection{CERN NA62}

\begin{figure}[!htb]
	\centering
	\includegraphics[width=0.9\linewidth]{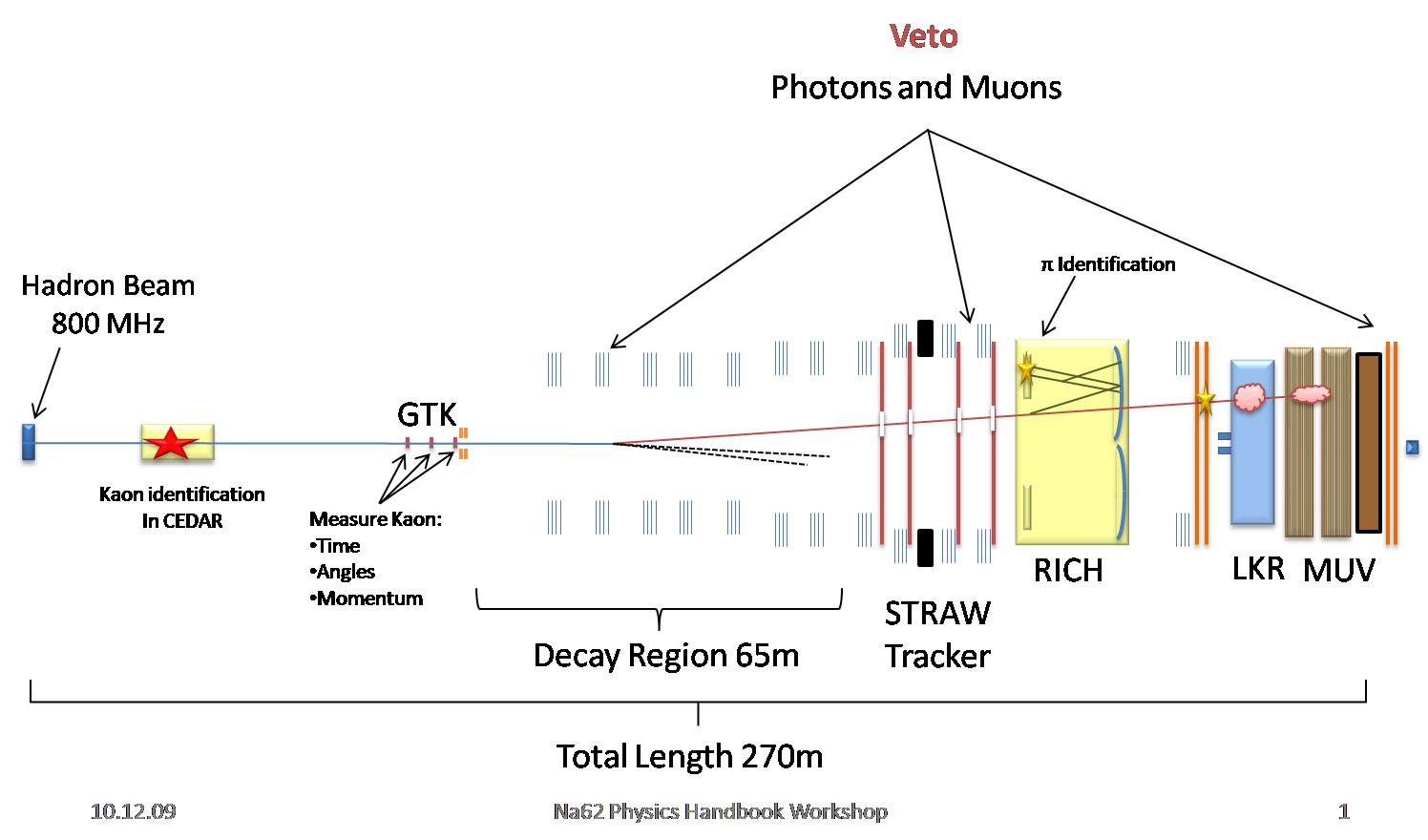}
	\caption{Schematic view of the CERN NA62 experimental apparatus~\cite{na62_url}.}
	\label{fig:na62_det}
\end{figure}

The CERN NA62 aims to measure the branching ratio of $\kppipnn$ 
by collecting 45 \sm\ events per year~\cite{na62_proposal}.
The experiment uses decay-in-flight $K^+$s instead of stopped $K^+$s
to run at a higher beam rate by eliminating a kaon-stopping-target
which would produce hadronic interactions.
A monochromatic 75-GeV/c $K^+$ beam enters the detector shown in Fig.~\ref{fig:na62_det}.
A kinematical cut is applied on 
the square of the missing mass
\((p_K - p_\pi)^2\), where $p_K$ and $p_\pi$ are kaon and pion 4-momenta, respectively.
The $p_K$ is measured with a silicon tracker located upstream of the decay region,
and the $p_\pi$ is measured with straw trackers and a spectrometer magnet downstream.
The $K^+$ and $\pi^+$ are identified with two
\v{C}erenkov counters (CEDAR and RICH in the figure).
The background from the $K^+ \to \pi^+\pi^0$ decay is suppressed 
by vetoing photons with lead glass photon veto rings on the side, and 
a liquid Kr calorimeter downstream.
The plan is to have an engineering run in late 2014, 
and start physics runs in 2015.
There is also a plan to search for lepton number violating decays such as 
\(K^+ \to \pi^+ \mu^\pm e^\mp\) and \(K^+ \to \pi^- \ell^+ \ell^+\) decays with a sensitivity of 
\(O(10^{-12})\).
Details of the NA62 was presented by A.~Romano at the CKM2014~\cite{romano}.

\subsection{J-PARC KOTO}

\begin{figure}[!htb]
	\centering
	\includegraphics[width=0.7\linewidth]{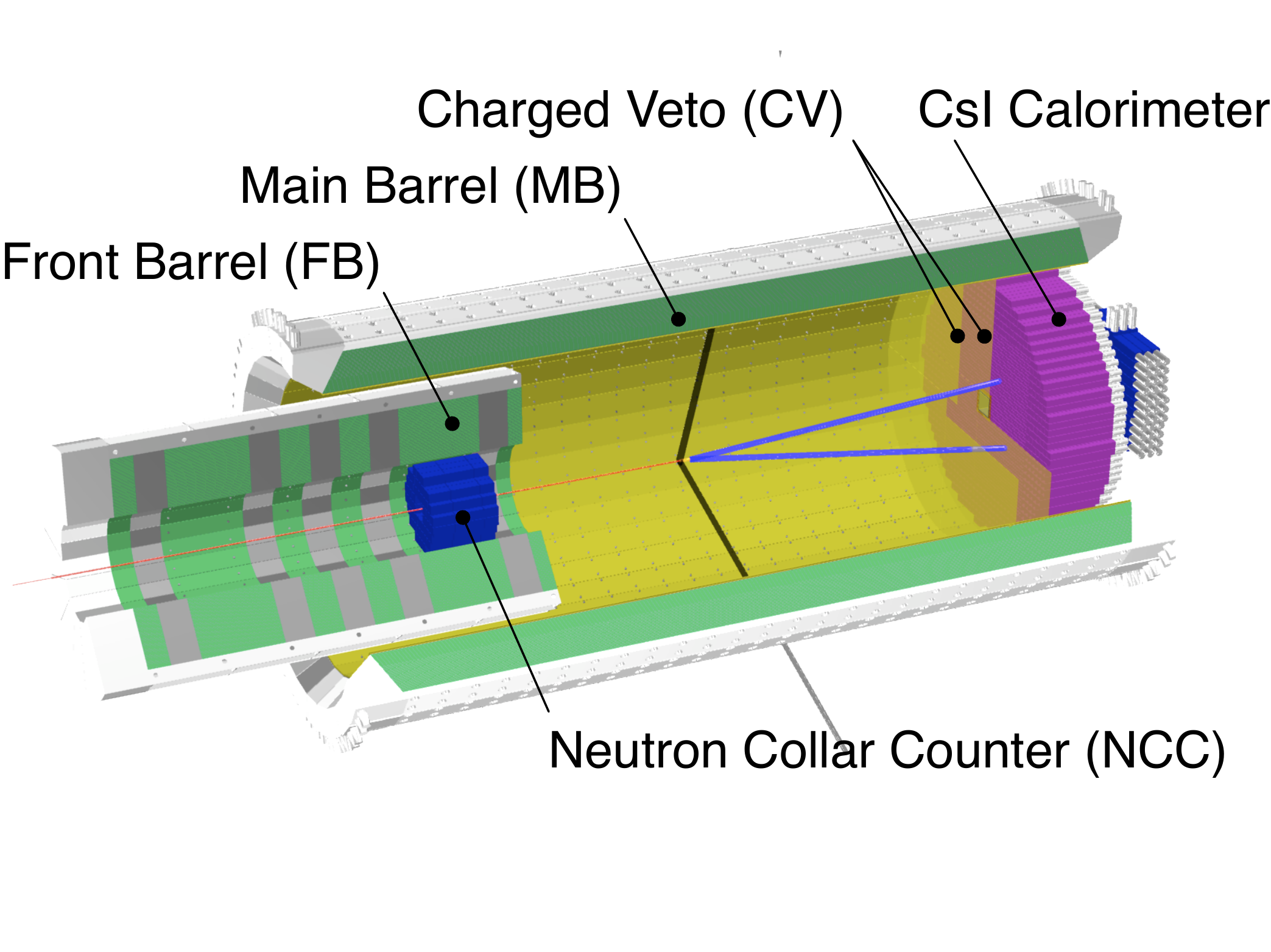}
	\caption{Schematic view of the J-PARC KOTO experimental apparatus.}
	\label{fig:koto_det}
\end{figure}

The J-PARC KOTO experiment is an experiment dedicated to search for the $\klpionn$ decay 
with a sensitivity close to the branching ratio predicted by the \sm~\cite{koto_proposal}.
It utilizes the high intensity 30-GeV proton beam at the J-PARC to produce a high flux of $K_L$s.
Figure~\ref{fig:koto_det} shows a schematic view of the KOTO detector.
The experiment searches for the decay by requiring two photons hitting 
the calorimeter placed at the downstream of a decay region, and that there are no other visible particles.
The major background comes from the $\klpiopio$ decay of which two of the four photons escape a detection.
To suppress the background, the decay region is hermetically covered by photon veto detectors.
To avoid dead materials such as a beam pipe between the neutral beam and the photon detectors, 
most of the detectors are placed inside a vacuum tank.
The calorimeter has a hole at the center to let the beam pass through.
Photons escaping through this hole are detected by modules consisting of a lead converter and 
an aerogel \v{C}erenkov counter.
The signals from all the detector channels are digitized every 8 ns or 2 ns to 
cope with their high counting rates.

The experiment took the first physics data in May 2013 for 100 hours.
The first result was presented by K.~Shiomi at the CKM2014 conference~\cite{shiomi}.

The plan of the experiment is to resume a run in 2015 and increase the statistics 
with a higher proton beam power.

\subsection{prospects}
\begin{figure}[!ht]
    \begin{center}
        \includegraphics[width=0.8\columnwidth]{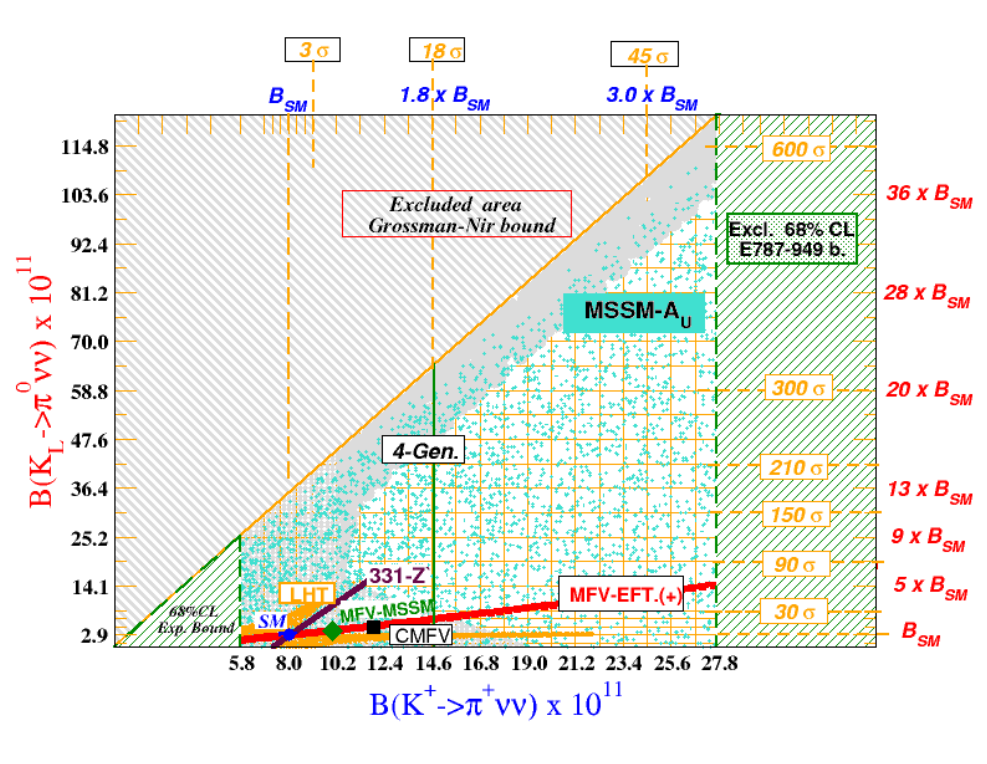}
        \caption{Branching ratios of $\klpionn$ vs $\kppipnn$ predicted by 
        		various theoretical models~\cite{mescia}.}
        \label{fig:kpinn_br}
    \end{center}
\end{figure}

Figure~\ref{fig:kpinn_br} shows the branching ratios of $\klpionn$ vs $\kppipnn$ 
predicted by various theoretical models~\cite{mescia}.
In a few years, most of the open area will be explored by the KOTO and NA62 experiments.

\section{Lepton Universality}
The \(\kenu\) and \(\kmunu\) decays can be used to test lepton universality.
The \sm\ predicts the ratio
\(R_K = BR(\kenu) / BR(\kmunu) = (2.477 \pm 0.001) \times 10^{-5}\)~\cite{cirigliano}.
The ratio can be different from the \sm\ prediction if there is a contribution from
charged Higgs, or charged Higgs plus a slepton.

\subsection{CERN NA62}
The CERN NA62 used a 74 GeV/c charged kaon beam, and collected the 
$\kenu$ and $\kmunu$ decays simultaneously.
It also had $K^-$ runs.
The decays were identified with the square of the missing mass
\((p_K - p_\ell)^2\), where $p_K$ and $p_\ell$ are kaon and lepton 4-momenta, respectively.
The leptons were identified from their $E/p$, 
the ratio between the energy deposit in the liquid Kr calorimeter,
and the measured track momentum.
The measured ratio was 
\(R_K = (2.488 \pm 0.007 \textrm{(stat.)} \pm 0.007 \textrm{(syst.)}) \times 10^{-5}\)~\cite{na62_RK}.

\subsection{J-PARC TREK-E36}
\begin{figure}[!htb]
	\centering
	\includegraphics[width=0.8\linewidth]{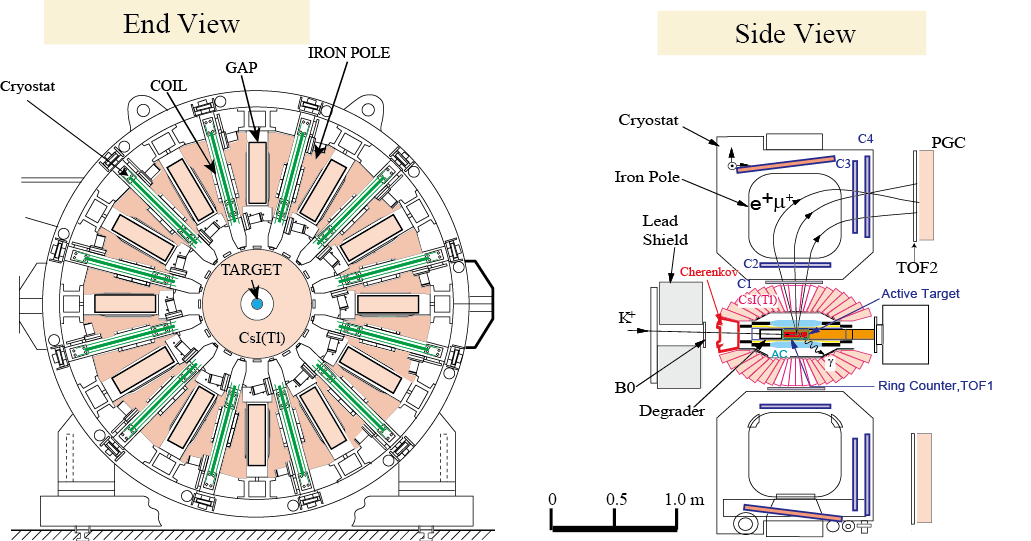}
	\caption{Schematic view of the J-PARC TREK-E36 experimental apparatus~\cite{e36_url}.}
	\label{fig:trek}
\end{figure}

The J-PARC TREK-E36 experiment is preparing to measure the $R_K$ with stopped $K^+$s.
As shown in Fig.~\ref{fig:trek}, the charged particles are momentum analyzed by a toroidal magnet, 
and photons are vetoed by CsI (Tl) crystals surrounding the target.
The experiment is planned to start running in 2015.
The expected errors are 0.20\% (stat.) and 0.15\% (syst.)~\cite{e36_url}.

\section{Other Kaon Experiments}
\subsection{KLOE-2}
KLOE-2 is an upgraded experiment that produces kaon pairs through
\(e^+e^- \to \phi \to K_S K_L, K^+K^-\).
It uses a crab-waist collision to increase the luminosity by a factor 3 compared to the previous KLOE experiment.
There are some detector upgrades to increase the acceptance and to improve the vertex resolution.
The plan is to collect 5 fb$^{-1}$ in the next 2--3 years.

\subsection{LHCb}
LHCb is known as a B-factory experiment, but it also works as a $K_S$ factory.
The detectors cover one side of the $p-p$ collision point
to detect particles boosted in the direction.
Because the experiment is designed to observe short-lived B mesons, 
it is also sensitive to $K_S$ decays.
The experiment recently gave $BR(K_S \to \mu^+\mu^-) < 9 \times 10^{-9}\) (90\% CL)~\cite{lhcb_ksmumu}.
Upgrades to the detectors and trigger are being planned to improve the limit to $O(10^{-10})$.
Studies of rare decays such as \(K_S \to e^+e^-\mu^+\mu^-\)
and \(e^+e^-e^+e^-\) are also being considered.


\section{Summary}
Kaon experiments are now focusing on searching for new physics beyond the \sm.
For example, 
CERN NA62, J-PARC KOTO and J-PARC TREK-E36 experiments are starting up 
to study $\kpinn$ decay modes, a lepton flavor violation, and lepton universality.



\Acknowledgements
I would like to thank the conference organizer for inviting me to give this talk.

\end{document}